\documentclass[letterpaper]{jpconf}
\usepackage{graphicx}
\usepackage{iopams}
\usepackage[latin1]{inputenc}
\usepackage{epsfig}

\newcommand{\text}[1]{\mbox{#1}}
\newcommand{\mathbbm}[1]{\mathbb{#1}}

\newcommand{\assign}{:=}
\newcommand{\bigintlim}{\int}
\newcommand{\bignone}{\,}
\newcommand{\mathpi}{\pi}
\newcommand{\mathd}{\mathrm{d}}
\newcommand{\mathe}{\mathrm{e}}
\newcommand{\mathi}{\mathrm{i}}
\newcommand{\tmmathbf}[1]{\ensuremath{\boldsymbol{#1}}}
\newcommand{\tmop}[1]{\ensuremath{\text{#1}}}

\newcommand{\emdash}{---}

\begin{document}

\title{Open quantum dynamics via environmental monitoring}
\author{Klaus Hornberger}
\address{Arnold Sommerfeld Center for Theoretical Physics,\\
 Ludwig-Maximilians-Universität München, \\
Theresienstraße 37, 80333 Munich, Germany}

\begin{abstract}
  A general method is discussed to obtain Markovian master equations which
  describe the interaction with the environment in a microscopic and
  non-perturbative fashion. It is based on combining time-dependent scattering
  theory with the concept of continuous quantum measurements. The applications
  to the case of a Brownian point particle and to the case of a complex
  molecule, both in the presence of a gaseous environment, are outlined.
\end{abstract}

\section{Introduction}

Quite a number of contributions presented at the DICE2006 workshop were
focused on possible extensions of standard quantum theory{\emdash}be it to
overcome the incompatibility of quantum theory and general relativity, or to
understand the measurement process and our perception of the world as
classical. When discussing these modifications of quantum theory, which often
imply a violation of unitarity, it is important to keep in mind that their
observability may be severely limited by environmental decoherence
{\cite{Joos2003a}}. The latter occurs naturally within the framework of
quantum theory if one accounts for the fact that all practically relevant
quantum systems are in contact with some uncontrollable environment. Before
interpreting the effects and judging the relevance of non-standard extensions
it is therefore of great importance to have a full quantitative understanding
of environmental decoherence.

In the present contribution I would like to discuss a new method to derive
Markovian master equations for open quantum systems, which take into account
the interaction with the environment in a {\emph{microscopic}} and
{\emph{non-perturbative}} fashion {\cite{Hornberger2006b,Hornberger2007b}}.
Such dynamic equations are a prerequisite for any faithful quantitative
description of open quantum systems. In particular, it is well-known that
phenomenological or perturbative master equations may fail by many orders of
magnitude when applied to situations where the decoherence time is much
shorter than the typical time scale of dissipation, even though they describe
dissipation phenomena rather well, see e.g. {\cite{Hornberger2007a}}.

In the standard approaches one starts out with an approximate total
Hamiltonian for the system plus environment {\cite{Breuer2002a}}. Then,
various approximations are employed in the course of the derivation, one of
which is often the Markov assumption. It implies that environmental
correlations disperse fast, so that on a coarse-grained timescale the temporal
change of the system state depends on the present state of the system, but not
on its history. The starting point of the method presented here differs
considerably from the standard methods. It is {\emph{not}} based on an
approximate ``total'' Hamiltonian of system plus environment. Rather, the
environmental coupling is described in an operational sense, by using
scattering theory, which permits a non-perturbative description. In
particular, this admits to take the Markov assumption as a premise, rather
than having to introduce it `by hand' in the course of the calculation.

The use and strength of the method will be exemplified by discussing the
dynamics of two different quantum systems in the presence of an ideal,
thermalized gas{\emdash}a trapped molecule and a Brownian point particle. The
corresponding master equations will be given in terms of the exact scattering
amplitudes describing the interaction of the gas particle with the quantum
system.

\section{The monitoring approach}

\begin{figure}[tb]
  \begin{center}
  \resizebox{13cm}{!}{\epsfig{file=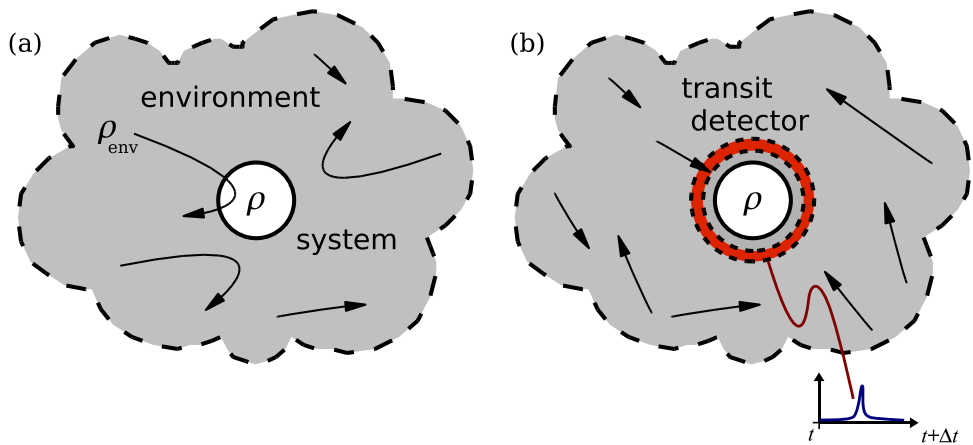}}
  \end{center}
  \caption{(a) It is assumed that the system interacts at most with one
  environmental \mbox{(quasi-)} particle at a time (no three-body collisions) and
  that they disperse their correlation with the system before scattering
  again. This is tantamount to the Markov assumption. (b) In order to
  consistently incorporate the state-dependence of the collision rate into the
  dynamic description, we imagine that the system is monitored continuously by
  a transit detector, which tells with a temporal resolution $\Delta t$
  whether a particle is going to scatter off the system, or
  not.\label{KHfig1}}
\end{figure}

The present approach to the derivation of master equations is motivated by the
observation that environmental decoherence, and Markovian open quantum
dynamics in general, can be viewed as due to the information transfer from the
system to the environment, see e.g.~{\cite{Hornberger2007a}}. In accordance
with this, we will picture the environment as consisting of (quasi-)particles,
which serve to monitor the system continuously in terms of microscopic probes
scattering off the system at random times. This point of view will be
applicable whenever one can describe the interaction with the environment in
terms of individual interaction events, and, of course, it suggests a
formulation in terms of scattering theory. The Markov assumption is then
easily incorporated from the outset by disregarding the change of the
environmental state after each collision.

When setting up a dynamic equation, one would like to write the change of the
system state in time as the infinitesimal rate of scatterings multiplied by
its effect, i.e., by the state transformation due to an individual collision.
However, not only the transformed state depends on the original system state,
in general, but also the collision rate, so that this naive ansatz would yield
a nonlinear equation. In order to account properly for the state dependence of
the collision rate we will make use of the concept of continuous and
generalized quantum measurements
{\cite{Kraus1983a,Busch1991a,Carmichael1993a}}. Specifically, and just for the
sake of consistent bookkeeping of the probabilities, we shall assume that the
system is surrounded by a hypothetical, minimally invasive transit detector,
which tells at any instant, and with a time resolution $\Delta t$, whether a
probe particle has passed by and is going to scatter off the system, see
Fig.~\ref{KHfig1}.

\subsection{The effect of continuous monitoring}

Let us denote by $C$ the event that the transit detector triggers in the
interval $\left( t ; t + \Delta t \right)$ indicating that a collision is
imminent. In general, the rate of collisions is then described by a positive
operator $\mathsf{\Gamma}$ acting in the the system-probe Hilbert space. Given
the uncorrelated state $\varrho_{\tmop{tot}} = \rho \otimes \rho_E$ it
determines the probability of a collision to occur in a small time interval
$\Delta t$,
\begin{eqnarray}
  \tmop{Prob} \left( C \text{ in $\Delta t$} | \rho \otimes \rho_E \right) & =
  & \Delta t \tmop{tr} \left( \mathsf{\Gamma} \left[ \rho \otimes \rho_E
  \right] \right) .  \label{KHeq:probscat}
\end{eqnarray}
Here $\rho_E$ is the stationary reduced single particle state of the
environment. The microscopic definition of $\mathsf{\Gamma}$ will in general
involve the current density operator of the relative motion and a total
scattering cross section, see below.

It is now important to note is that the information that a collision is going
to take place changes our knowledge about the state, and therefore its
description. Most generally, it is transformed by a generalized measurement
transformation, i.e., a trace decreasing, completely positive map
{\cite{Kraus1983a,Busch1991a}}. At the same time, we have to take into
consideration that the detector is not real, but is introduced here only for
enabling us to account for the state dependence of the collision probability.
It is therefore reasonable to take the detection process as {\emph{efficient}}
and {\emph{minimally-invasive}}, so that it is described by a single operator
which introduces no reversible back-action. The consistency requirement for
measurement transformations then implies that after a (hypothetical) detector
click, but prior to scattering, the normalized system-probe state must have
the form
\begin{eqnarray}
  \mathcal{M} \left( \varrho_{\tmop{tot}} |C \right) & = & \frac{
  \mathsf{\Gamma}^{1 / 2} \varrho_{\tmop{tot}}  \mathsf{\Gamma}^{1 /
  2}}{\tmop{tr} \left( \mathsf{\Gamma} \varrho_{\tmop{tot}} \right)} . 
  \label{KHeq:mestra}
\end{eqnarray}
This nonlinear transformation reflects our improved knowledge about the
incoming two-particle wave packet, and it may be viewed as enhancing those
parts which are heading towards a collision. Also the absence of a detection
event during $\Delta t$ constitutes a measurement which changes the state. The
corresponding probability must satisfy ${\tmop{Prob} \left( \bar{C} \text{ in
$\Delta t$} | \varrho_{\tmop{tot}} \right)} = 1 - {\tmop{Prob} \left( C \text{
in $\Delta t$} | \varrho_{\tmop{tot}} \right)}$, and the state conditioned on
a null-event is given by
\begin{eqnarray}
  \mathcal{M} \left( \varrho_{\tmop{tot}} | \bar{C} \right) & = &
  \frac{\varrho_{\tmop{tot}} - \Delta t \mathsf{\Gamma}^{1 / 2}
  \varrho_{\tmop{tot}}  \mathsf{\Gamma}^{1 / 2}}{1 - \Delta t \tmop{tr} \left(
  \mathsf{\Gamma} \varrho_{\tmop{tot}} \right)} .  \label{KHeq:compmestra}
\end{eqnarray}
In fact, more general nonlinear transformations are conceivable, but this one
is distinguished by the fact that it introduces no further operators.

The effect of a single collision is described by the two-particle S-matrix
$\mathsf{S}$. The unconditioned system-probe state after time $\Delta t$ can
now be formed by taking into account that the detection outcomes are not
really available. The infinitesimally evolved state is then given by the sum
of the colliding state transformed by the S-matrix and the untransformed
non-colliding one, weighted with their respective probabilities,
\begin{eqnarray}
  \varrho'_{\tmop{tot}} \left( \Delta t \right) & = & \tmop{Prob} \left( C
  \text{ in $\Delta t$} | \varrho_{\tmop{tot}} \right) \mathsf{S} \mathcal{M}
  \left( \varrho_{\tmop{tot}} |C \right) \mathsf{S}^{\dag} + \tmop{Prob}
  \left( \bar{C} \text{ in $\Delta t$} | \varrho_{\tmop{tot}} \right)
  \mathcal{M} \left( \varrho_{\tmop{tot}} | \bar{C} \right) \nonumber\\
  & = & \mathsf{S} \mathsf{\Gamma}^{1 / 2} \varrho_{\tmop{tot}} 
  \mathsf{\Gamma}^{1 / 2} \mathsf{S}^{\dag} \Delta t + \varrho_{\tmop{tot}} -
  \mathsf{\Gamma}^{1 / 2} \varrho_{\tmop{tot}}  \mathsf{\Gamma}^{1 /
    2} \Delta  t.  \label{KHeq:rhoinf}
\end{eqnarray}

\subsection{Constructing a dynamic equation}

In order to obtain a differential equation for the time evolution it is now
convenient to split off the nontrivial part of the two-particle S-matrix, thus
defining
\begin{eqnarray}
  \mathsf{T} & = & \mathi \left( \mathbbm{I} - \mathsf{S} \right) . 
\end{eqnarray}
This operator is proportional to the $T$-matrix of scattering theory (only) on
the energy shell, and one finds that the unitarity of $\mathsf{S}$ implies
that it satisfies $\mathi \left( \mathsf{T} - \mathsf{T}^{\dag} \right) = -
\mathsf{T}^{\dag} \mathsf{T}$. Using this relation the differential quotient
can be written as
\begin{eqnarray}
  \frac{\varrho' \left( \Delta t \right) - \varrho}{\Delta t} & = &
  \mathsf{T} \mathsf{\Gamma}^{1 / 2} \varrho \mathsf{\Gamma}^{1 / 2}
  \mathsf{T}^{\dag} - \frac{1}{2} \mathsf{T}^{\dag} \mathsf{T}
  \mathsf{\Gamma}^{1 / 2} \varrho \mathsf{\Gamma}^{1 / 2} - \frac{1}{2}
  \mathsf{\Gamma}^{1 / 2} \varrho \mathsf{\Gamma}^{1 / 2} \mathsf{T}^{\dag}
  \mathsf{T} + \frac{\mathi}{2} \left[ \mathsf{T} + \mathsf{T}^{\dag},
  \mathsf{\Gamma}^{1 / 2} \varrho \mathsf{\Gamma}^{1 / 2} \right] . 
\end{eqnarray}
From here it is easy to obtain a closed differential equation for $\rho$. We
trace out the environment with $\varrho = \rho \otimes \rho_E$, take the limit
of continuous monitoring $\Delta t \rightarrow 0$, to arrive at
{\cite{Hornberger2007b}}
\begin{eqnarray}
  \frac{\mathd}{\mathd t} \rho & = & \frac{1}{\mathi \hbar} \left[ \mathsf{H},
  \rho \right] + \frac{\mathi}{2} \tmop{Tr}_E \left( \left[ \mathsf{T} +
  \mathsf{T}^{\dag}, \mathsf{\Gamma}^{1 / 2} \left[ \rho \otimes \rho_E
  \right] \mathsf{\Gamma}^{1 / 2} \right] \right) + \tmop{Tr}_E \left(
  \mathsf{T} \mathsf{\Gamma}^{1 / 2} \left[ \rho \otimes \rho_E \right]
  \mathsf{\Gamma}^{1 / 2} \mathsf{T}^{\dag} \right) \nonumber\\
  &  & - \frac{1}{2} \tmop{Tr}_E \left( \mathsf{\Gamma}^{1 / 2}
  \mathsf{T}^{\dag} \mathsf{T}  \mathsf{\Gamma}^{1 / 2} \left[ \rho \otimes
  \rho_E \right] \right) - \frac{1}{2} \tmop{Tr}_E \left( \left[ \rho \otimes
  \rho_E \right]  \mathsf{\Gamma}^{1 / 2} \mathsf{T}^{\dag} \mathsf{T}
  \mathsf{\Gamma}^{1 / 2} \right) .  \label{KHeq:me1}
\end{eqnarray}
The Markov approximation enters here by assuming the factorization $\rho
\otimes \rho_E$ to be valid for all times. Note also that the generator
$\mathsf{H}$ of the free system evolution was added, thus switching from the
interaction picture to the Schr\"odinger picture.
The collision rate with its state dependence is
incorporated by the operators $\mathsf{\Gamma}^{1 / 2}$, while the operators
$\mathsf{T}$ describe the individual microscopic interaction process without
approximation.

The discussion was very general, so far. However, to obtain concrete master
equations system and environment have to be specified, along with the
operators $\mathsf{\Gamma}$ and $\mathsf{S}$ describing their interaction. In
the following applications, we will assume the environment to be an ideal
Maxwell gas. Its single particle state
\begin{eqnarray}
  \rho_{\tmop{gas}} & = & \frac{\Lambda_{\tmop{th}}^3}{\Omega} \exp \left( -
  \beta \frac{\mathsf{p}^2}{2 m} \right)  \label{KHeq:gas}
\end{eqnarray}
is characterized by the inverse temperature $\beta$, the normalization volume
$\Omega$, and the thermal de Broglie wave length $\Lambda_{\tmop{th}} =
\sqrt{2 \mathpi \hbar^2 \beta / m}$.

\section{Collisional decoherence of a discrete object}

\begin{figure}[tb]
  \resizebox{10cm}{!}{\epsfig{file=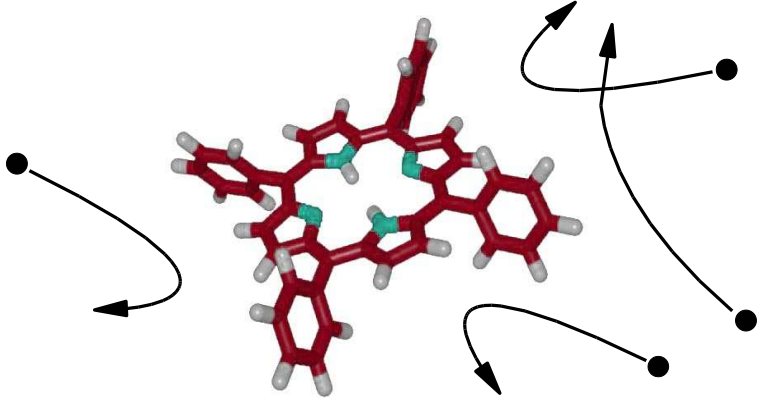}}
  \hspace{3mm}
  \begin{minipage}[b]{5.7cm}
  \caption{Superpositions of different configuration states of a molecule will
  decohere into a mixture of configuration states, as a result of the
  interaction with a gaseous environment. The master equation
  (\ref{KHeq:master}) describes this process in a non-perturbative fashion.
  \label{fig2}}
  \end{minipage}
\end{figure}

As a first concrete implementation of the monitoring approach, let us see how
an immobile object with discrete internal states, such as the electronic,
configuration, and rotation dynamics of a trapped molecule, gets affected by
an environment of ideal gas particles. For example, such a master equation
will describe how a possible superposition of configuration eigenstates of
macromolecules decoheres into a mixture of robust configuration states, see
Fig.~\ref{fig2}.

The interaction between system and gas will be described in terms of the exact
scattering amplitudes $f_{\alpha \alpha_0} \left( \tmmathbf{p}, \tmmathbf{p}_0
\right)$ determined by the interaction potential. They describe the in
general inelastic coupling between the internal energy eigenstates, which form
a discrete basis of the system Hilbert space and are called {\emph{channels}}
in the language of scattering theory. The notation $| \alpha \rangle$ will be
used in the following to indicate the system eigenstates of energy
$E_{\alpha}$. In this channel basis, $\rho_{\alpha \beta} = \langle \alpha |
\rho | \beta \rangle$, the equation of motion (\ref{KHeq:me1}) takes on the
form of a general discrete master equation of Lindblad type,
\begin{eqnarray}
  \partial_t \rho_{\alpha \beta} & = & - \mathi \omega_{\alpha \beta}
  \rho_{\alpha \beta} + \sum_{\alpha_0 \beta_0} \rho_{\alpha_0 \beta_0}
  \bignone M_{\alpha \beta}^{\alpha_0 \beta_0} - \frac{1}{2}  \sum_{\alpha_0
  \gamma} \rho_{\alpha_0 \beta_{}} \bignone \bignone M_{\gamma
  \gamma}^{\alpha_0 \alpha} - \frac{1}{2}  \sum_{\beta_0 \gamma} \rho_{\alpha
  \beta_0} \bignone M_{\gamma \gamma}^{\beta \beta_0} .  \label{KHeq:master}
\end{eqnarray}
Here, the frequencies $\omega_{\alpha \beta} = \left[ E_{\alpha} - E_{\beta} +
\left( \varepsilon_{\alpha} - \varepsilon_{\beta} \right) \right] / \hbar$
involve energy shifts $\varepsilon_{\alpha} \in \mathbbm{R}$, which are
defined below. They describe the coherent modification of the system energies
due to the presence of the environment analogous to the Lamb shift. More
importantly, the incoherent dynamics effected by the environment is described
by the set of complex rate coefficients
\begin{eqnarray}
  M_{\alpha \beta}^{\alpha_0 \beta_0} & = & \langle \alpha | \tmop{Tr}_E
  \left( \mathsf{T} \mathsf{\Gamma}^{1 / 2} \left[ | \alpha_0 \rangle \langle
  \beta_0 | \otimes \rho_{\tmop{gas}} \right] \mathsf{\Gamma}^{1 / 2}
  \mathsf{T}^{\dag} \right) | \beta \rangle .  \label{KHeq:M}
\end{eqnarray}
Our main task is to calculate these quantities. As a fist step we need to
specify the rate operator $\mathsf{\Gamma}$. It is naturally given in terms of
the current density operator $\mathsf{j} = n_{\tmop{gas}}  \mathsf{p} / m$ of
the impinging gas particles multiplied by the channel-specific total
scattering cross sections $\sigma \left( \tmmathbf{p}, \alpha \right)$,
\begin{eqnarray}
  \mathsf{\Gamma} & = & n_{\tmop{gas}} \sum_{\alpha} \mathsf{\Pi}_{\alpha}
  \otimes \frac{\left| \mathsf{p} \right|}{m} \sigma \left( \mathsf{p}, \alpha
  \right),  \label{KHeq:Gamma}
\end{eqnarray}
with projectors $\mathsf{\mathsf{\Pi}}_{\alpha} = | \alpha \rangle \langle
\alpha |$.

For normalized and separable system-gas states the expectation value of
$\mathsf{\Gamma}$ thus yields the infinitesimal total collision
probability{\emdash}provided the motional state of the gas particle is a wave
packet heading towards the origin. However, since (\ref{KHeq:Gamma}) depends
only on the modulus of the velocity $\left| \mathsf{p} \right| / m$ it will
yield a finite collision probability even if the particle is heading away from
the origin. Hence, for (\ref{KHeq:me1}) to make sense either the definition of
$\mathsf{\Gamma}$ should contain in addition a projection to the subset of
incoming states, or the S-matrix should keep such a non-colliding state
unaffected. The latter is not guaranteed in general, since even a purely
outgoing wave packet, located far away form the origin, may get transformed by
$\mathsf{S}$ since the definition of the S-matrix involves a backward
time-evolution {\cite{Taylor1972a}}.

\subsection{A phase space-based calculation}

In the following, we will explicitly disregard the outgoing states by using a
non-diagonal decomposition of $\rho_{\tmop{gas}}$. As discussed in
{\cite{Hornberger2003b}} the thermal gas state can be written as a phase space
integration over projectors onto Gaussian states of minimum uncertainty,
\begin{eqnarray}
  | \psi_{\tmmathbf{r}_0 \tmmathbf{p}_0} \rangle & = & \left( \frac{2 \pi
  \hbar^2 \bar{\beta}}{m} \right)^{3 / 4} \exp \left( - \bar{\beta}
  \frac{\left( \mathsf{p} -\tmmathbf{p}_0 \right)^2}{4 m} \right)
  |\tmmathbf{r}_0 \rangle 
\end{eqnarray}
characterized by an inverse temperature $\bar{\beta}$ greater than the inverse
gas temperature $\beta$. Denoting the Maxwell-Boltzmann distribution
corresponding to the temperature $\hat{\beta}^{- 1} = \beta^{- 1} -
\bar{\beta}^{- 1}$ by $\hat{\mu} \left( \tmmathbf{p}_0 \right) = (2 \pi m /
\hat{\beta})^{- 3 / 2} \exp (- \hat{\beta} \tmmathbf{p}_0^2 / 2 m)$ the state
(\ref{KHeq:gas}) is represented as
\begin{eqnarray}
  \rho_{\tmop{gas}} & = & \bignone \int \mathd \tmmathbf{p}_0  \hat{\mu}
  \left( \tmmathbf{p}_0 \right) \int_{\Omega} \frac{\mathd
  \tmmathbf{r}_0}{\Omega} \bignone | \psi_{\tmmathbf{r}_0 \tmmathbf{p}_0}
  \rangle \langle \psi_{\tmmathbf{r}_0 \tmmathbf{p}_0} |.  \label{eq:rhops}
\end{eqnarray}
Inserting (\ref{eq:rhops}) into (\ref{KHeq:M}) yields the expression
\begin{eqnarray}
  M_{\alpha \beta}^{\alpha_0 \beta_0} & = & \int \mathd \tmmathbf{p}_0 
  \hat{\mu} \left( \tmmathbf{p}_0 \right) \int_{\Omega} \frac{\mathd
  \tmmathbf{r}_0}{\Omega} \bignone m_{\alpha \beta}^{\alpha_0 \beta_0} \left(
  \tmmathbf{r}_0, \tmmathbf{p}_0 \right)  \label{eq:M1}
\end{eqnarray}
with
\begin{eqnarray}
  m_{\alpha \beta}^{\alpha_0 \beta_0} \left( \tmmathbf{r}_0, \tmmathbf{p}_0
  \right) & \assign & \int \mathd \tmmathbf{p} \bignone \langle \alpha |
  \langle \tmmathbf{p}| \mathsf{T} \Gamma_{}^{1 / 2} | \alpha_0 \rangle |
  \psi_{\tmmathbf{r}_0 \tmmathbf{p}_0} \rangle \langle \beta_0 | \langle
  \psi_{\tmmathbf{r}_0 \tmmathbf{p}_0} | \Gamma_{}^{1 / 2}  \mathsf{T^{\dag}}
  | \beta \rangle |\tmmathbf{p} \rangle .  \label{eq:m1}
\end{eqnarray}
Evidently, the phase space function $m_{\alpha \beta}^{\alpha_0 \beta_0}
\left( \tmmathbf{r}_0, \tmmathbf{p}_0 \right)$ gives the contribution of
different phase space regions to the rate coefficient $M_{\alpha
\beta}^{\alpha_0 \beta_0}$.

This permits now to restrict the calculation to incoming wave packets. Since
the $m_{\alpha \beta}^{\alpha_0 \beta_0}$ are averaged over all available
positions in (\ref{eq:M1}) it is natural to confine this spatial average at
fixed $\tmmathbf{p}_0$ to a cylinder pointing in the direction of
$\tmmathbf{p}_0$, whose longitudinal support $\Lambda_{\tmmathbf{p}_0}$
vanishes at outgoing positions. Its transverse base area is given by an
average cross section $\Sigma_{\tmmathbf{p}_0}$. In order to evaluate the
phase space function (\ref{eq:m1}), one inserts momentum resolutions of unity
between the $\mathsf{T}$ and $\Gamma$ operators and notes the representation
{\cite{Taylor1972a}}
\begin{eqnarray}
  \langle \alpha_f | \langle \tmmathbf{p}_f | \mathsf{T} | \alpha_i \rangle
  |\tmmathbf{p} \text{$_i$} \rangle & = &  \frac{f_{\alpha_f \alpha_i} \left(
  \tmmathbf{p}_f, \tmmathbf{p}_i \right)}{2 \pi \hbar m} \delta \left( E_{p_f
  \alpha_f} - E_{p_i \alpha_i} \right)  \label{eq:Trep}
\end{eqnarray}
in terms of the multi-channel scattering amplitude and total energies $E_{p
\alpha_{}} = p^2 / 2 m + E_{\alpha}$. In the calculation one chooses
$\bar{\beta}$ large and eventually takes the limit $\bar{\beta} \rightarrow
\infty$, $\hat{\beta} \rightarrow \beta$ of very extended wave packets so that
$\hat{\mu}$ approaches the original Maxwell-Boltzmann distribution $\mu$. In
this limit the dependence on $\Lambda_{\tmmathbf{p}_0}$ and
$\Sigma_{\tmmathbf{p}_0}$ drops out provided one identifies
$\Sigma_{\tmmathbf{p}_0}$ with the geometric mean of the total cross sections
of the involved channels. One obtains {\cite{Hornberger2007b}}
\begin{eqnarray}
  M_{\alpha \beta}^{\alpha_0 \beta_0} & = & \chi_{\alpha \beta}^{\alpha_0
  \beta_0}  \frac{n_{\tmop{gas}}}{m^2} \int \mathd \tmmathbf{p} \bignone
  \bignone \mathd \tmmathbf{p}_0 \mu \left( \tmmathbf{p}_0 \right) f_{\alpha
  \alpha_0} \left( \tmmathbf{p}, \tmmathbf{p}_0  \right) f_{\beta
  \beta_0}^{\ast} \left( \tmmathbf{p}, \tmmathbf{p}_0 \right) \delta \left(
  \frac{\tmmathbf{p}^2 -\tmmathbf{p}_0^2}{2 m} + E_{\alpha} - E_{\alpha_0}
  \right) \nonumber\\
  &  &  \label{KHeq:M3}
\end{eqnarray}
with the Kronecker-like factor $\chi_{\alpha \beta}^{\alpha_0 \beta_0}$, which
is equal to one if $E_{\alpha} - E_{\alpha_0} = E_{\beta} - E_{\beta_0}$ and
zero otherwise. Moreover, the energy shifts are determined the real parts of
the forward scattering amplitude,
\begin{eqnarray}
  \varepsilon_{\alpha} & = & - 2 \pi \hbar^2 \frac{n_{\tmop{gas}}}{m} \int
  \mathd \tmmathbf{p}_0 \mu \left( \tmmathbf{p}_0 \right) \tmop{Re} \left[
  f_{\alpha \alpha} \left( \tmmathbf{p}_0, \tmmathbf{p}_0 \right) \right] . 
\end{eqnarray}
For the special case of factorizing interactions, $\mathsf{H}_{\tmop{int}} =
\mathsf{A} \otimes \mathsf{B}_E$, and for times large compared to all system
time scales this result can be obtained rigorously {\cite{Dumcke1985a}}, by
means of a ``low density limit'' scaling method
{\cite{Alicki1987a,Breuer2002a}}. As discussed in {\cite{Hornberger2007b}},
limiting forms of (\ref{KHeq:master}) display the expected dynamics. On the
diagonal, it reduces to a rate equation, where the total inelastic cross
sections determine the transition rates, while in the case of elastic
scattering the coherences decay exponentially.

\subsection{An equivalent approach}

It is worth noting that the result (\ref{KHeq:M3}) for the rate coefficients
$M_{\alpha \beta}^{\alpha_0 \beta_0}$ can as well be obtained in a more
direct, while less obvious way, if the diagonal momentum representation of
$\rho_{\tmop{env}}$ is used instead of (\ref{eq:rhops}). In this case, the
improper eigenstates $|\tmmathbf{p} \rangle$ have a meaning beyond their use
as convenient basis states for the expansion of the incoming state. Rather,
each momentum state in $\rho_{\tmop{env}}$ must be viewed as representing the
limiting form of a {\emph{normalized}} wave packet. It is therefore
necessarily restricted to a discrete set defined by the corresponding
volume-normalized states $| \tilde{\tmmathbf{p}} \rangle$. The problem is then
that the application of $\mathsf{S}$ to these spatially extended states leads
to the unwanted transformation also of its ``outgoing components''. As a
consequence, the resulting expression for $M_{\alpha \beta}^{\alpha_0
\beta_0}$ involving
\begin{eqnarray}
  &  & \frac{\left( 2 \pi \hbar \right)^3}{\Omega} \langle \alpha
  \tmmathbf{p}_1 | \mathsf{T}_0 | \alpha_0 \tmmathbf{p}_0 \rangle \langle
  \beta_0 \tmmathbf{p}_0 | \mathsf{T}_0^{\dag} | \beta \tmmathbf{p}_1 \rangle 
  \label{eq:ww1}
\end{eqnarray}
is ill-defined since it contains the square of the $\delta$-functions in
(\ref{eq:Trep}) and the normalization volume $\Omega$.

This can be healed by the second option mentioned above, the formal
modification of the operator $\mathsf{S}$, such that it keeps outgoing
components invariant. In the continuous basis of improper states the unitarity
of $\mathsf{S}$ is expressed by the optical theorem, which quantifies the
diffraction limitation of the scattering probability. Since diffraction cannot
be accommodated in the discrete basis, the requirement of probability current
conservation must be incorporated additionally in any consistent modification
of $\mathsf{S}$, which serves to define a normalized transition matrix for the
``scattering'' into the different ``leads'' characterized by the discrete $|
\tilde{\tmmathbf{p}} \rangle$. The point is now that this normalization
condition can again be viewed in the continuous basis of improper states, and
thus provides a simple rule how to form a well-defined expression
{\cite{Hornberger2003b}}. Its multichannel version is to replace
(\ref{eq:ww1}) by
\begin{eqnarray}
  &  &  \frac{\chi_{\alpha \beta}^{\alpha_0 \beta_0}}{p_0 m} \frac{f_{\alpha
  \alpha_0} \left( \tmmathbf{p}, \tmmathbf{p}_0 \right) f^{\ast}_{\beta
  \beta_0} \left( \tmmathbf{p}, \tmmathbf{p}_0 \right)}{\sqrt{\sigma \left(
  p_0, \alpha_0 \right) \sigma \left( p_0, \beta_0 \right)} } \delta \left(
  \frac{\tmmathbf{p}^2 -\tmmathbf{p}_0^2}{2 m} + E_{\alpha} - E_{\alpha_0}
  \right) .  \label{eq:ww2}
\end{eqnarray}
The appearance of the cross sections in the denominator is here a direct
consequence of the normalization requirement for the probability current. By
using this replacement one obtains the result (\ref{KHeq:M3}) immediately from
(\ref{KHeq:M}) for any momentum diagonal $\rho_{\tmop{env}}$.

It should be mentioned that the problem of squared delta functions can also be
circumvented, as discussed in {\cite{Diosi1995a,Adler2006a}}, by appealing to
the Fermi golden rule when one writes down the dynamic equation. This way one
leaves the realm of (time-dependent) scattering theory {\cite{Taylor1972a}},
since one needs to speak about the ``elapsed time'' during the scattering
process. Effectively, the ``interaction Hamiltonian'' of the dynamic
description is thus identified with the off-shell T-matrix, a step which is
valid (only) in second-order perturbation theory. It should be noted that the
form of such a Born approximation to the master equation can differ from the
non-perturbative result, as demonstrated in the following section.

\section{Collisional decoherence of a Brownian particle}

As a second, nontrivial application of the monitoring approach, let us
consider how the motional state of a single Brownian particle is affected by a
gaseous environment. In analogy to the classical case, the corresponding
master equation may be called a linear quantum Boltzmann equation. Here the
term ``linear'' refers to the fact that the equation for the single Brownian
particle is linear (the particle is not interacting with itself), and should
not be confused with the ``linearized'', i.e., perturbative description of the
reduced single particle state a self-interacting gas. The resulting equation
will describe, on equal footing, both the short-time decoherence behavior,
i.e. the rapid ``localization'' of a spatial superposition state into a
mixture, and the long-time dissipative behavior, i.e. the gradual
thermalization of the Brownian particle.

Let us denote by $M$ and $m$ the masses of the Brownian and the gas particle,
and the reduced mass by $m_{\ast}$. Since a collision affects only the
relative coordinates we denote the single particle S-matrix by $\mathsf{S}_0 =
\mathbbm{I} + i \mathsf{T}_0$. Moreover, it is convenient to denote the
relative momentum by
\begin{eqnarray}
  \tmop{rel} \left( \tmmathbf{p}, \tmmathbf{P} \right) & \assign &
  \frac{m_{\ast}}{m} \tmmathbf{p}- \frac{m_{\ast}}{M} \tmmathbf{P}. 
\end{eqnarray}
Like above, the first step in obtaining the master equation is to specify the
rate operator. In classical mechanics, the rate is determined by the modulus
of the relative current density $j_{\tmop{rel}} \left( \tmmathbf{p},
\tmmathbf{P} \right) = n_{\tmop{gas}}  \left| \tmop{rel} \left( \tmmathbf{p},
\tmmathbf{P} \right) \right| / m_{\ast}$ multiplied by the total scattering
cross section $\sigma \left( \tmmathbf{p}_{\tmop{in}} \right)$. This suggests
\begin{eqnarray}
  \mathsf{\Gamma} & = & \bignone j_{\tmop{rel}} \left( \mathsf{p}, \mathsf{P}
  \right) \sigma \left( \tmop{rel} \left( \mathsf{p}, \mathsf{P} \right)
  \right),  \label{eq:Gam3}
\end{eqnarray}
and, indeed, for normalized and separable particle-gas states the expectation
value of this operator yields the collision rate experienced by the Brownian
particle. Like above, the definition (\ref{eq:Gam3}) does not yet include the
projection to the subspace of incoming particle-gas wave packets.

\subsection{Calculation in the momentum diagonal basis}

Inserting the diagonal momentum representation of the gas state
(\ref{KHeq:gas}) into (\ref{KHeq:me1}) we find that the equation of motion for
the Brownian particle in momentum representation, $\rho \left( \tmmathbf{P},
\tmmathbf{P}' \right) = \langle \tmmathbf{P}| \rho |\tmmathbf{P}' \rangle$,
takes the form
\begin{eqnarray}
  \partial_t \rho \left( \tmmathbf{P}, \tmmathbf{P}' \right) & = & \int \mathd
  \tmmathbf{Q} \bignone M_{\tmop{in}} \left( \tmmathbf{P}, \tmmathbf{P}' ;
  \tmmathbf{Q} \right) \rho \left( \tmmathbf{P}-\tmmathbf{Q}, \tmmathbf{P}'
  -\tmmathbf{Q} \right) \nonumber\\
  &  & - \frac{1}{2} \left[ M_{\tmop{out}}^{\tmop{cl}} \left( \tmmathbf{P}
  \right) + M_{\tmop{out}}^{\tmop{cl}} \left( \tmmathbf{P}' \right) \right]
  \rho \left( \tmmathbf{P}, \tmmathbf{P}' \right) .  \label{eq:Dtrho2}
\end{eqnarray}
In particular, there is no contribution of the second term in
(\ref{KHeq:me1}), since the gas density is uniform. Equation (\ref{eq:Dtrho2})
is specified by the complex rate function {\cite{Hornberger2006b}}
\begin{eqnarray}
  M_{\tmop{in}} \left( \tmmathbf{P}, \tmmathbf{P}' ; \tmmathbf{Q} \right) & =
  & \frac{n_{\tmop{gas}}}{m_{\ast}}  \int \mathd \tmmathbf{p}_0 \mu \left(
  \tmmathbf{p}_0 \right)  \sqrt{\left| \tmmathbf{p}_i +\tmmathbf{q} \right|
  \sigma \left( \tmmathbf{p}_i +\tmmathbf{q} \right)}  \sqrt{\left|
  \tmmathbf{p}_i -\tmmathbf{q} \right| \sigma \left( \tmmathbf{p}_i
  -\tmmathbf{q} \right)} \nonumber\\
  &  & \times \frac{\left( 2 \mathpi \hbar \right)^3}{\Omega} \langle
  \tmmathbf{p}_f +\tmmathbf{q}| \mathsf{T}_0 |\tmmathbf{p}_i +\tmmathbf{q}
  \rangle \langle \tmmathbf{p}_i -\tmmathbf{q}| \mathsf{T}_0^{\dag}
  |\tmmathbf{p}_f -\tmmathbf{q} \rangle .  \label{eq:Min11}
\end{eqnarray}
Here $\tmmathbf{p}_i \assign \tmop{rel} \left( \tmmathbf{p}_0,
\frac{\tmmathbf{P}+\tmmathbf{P}'}{2} -\tmmathbf{Q} \right)$ and
$\tmmathbf{p}_f \assign \tmmathbf{p}_i -\tmmathbf{Q}$ were introduced as
functions of $\tmmathbf{p}_0$. Moreover, \ {$\tmmathbf{q} \assign \tmop{rel}
\left( 0, \frac{\tmmathbf{P}-\tmmathbf{P}'}{2} \right)$}. The rate function
(\ref{eq:Min11}) defines also
\begin{eqnarray}
  M_{\tmop{out}}^{\tmop{cl}} \left( \tmmathbf{P} \right) & = & \bigintlim
  \mathd \tmmathbf{Q} \, M_{\tmop{in}} \left( \tmmathbf{P}+\tmmathbf{Q},
  \tmmathbf{P}+\tmmathbf{Q}; \tmmathbf{Q} \right), 
\end{eqnarray}
which will turn out to be the rate in the classical linear Boltzmann equation
of a particle with momentum $\tmmathbf{P}$ to be scattered by the gas into a
different direction or velocity. The remaining difficulty is to evaluate the
function (\ref{eq:Min11}). Like in the previous section, our use of a diagonal
representation of the normalized gas state leads to an ill-defined expression
involving a square of delta-functions and the normalization volume, which can
be traced back to the unwanted transformation of the outgoing parts of the
normalized state,
\begin{eqnarray}
  &  & \frac{\left( 2 \pi \hbar \right)^3}{\Omega} \langle \tmmathbf{p}_f
  +\tmmathbf{q}| \mathsf{T}_0 | \tmmathbf{p_i} +\tmmathbf{q} \rangle \langle
  \tmmathbf{p}_i -\tmmathbf{q}| \mathsf{T}_0^{\dag} |\tmmathbf{p}_f
  -\tmmathbf{q} \rangle g \left( \tmmathbf{q} \right) .  \label{eq:rr1}
\end{eqnarray}
Like above, it can be argued that any consistent modification of the S-matrix,
i.e., of $\mathsf{T}_0$, which keeps the outgoing parts invariant must
necessarily transform (\ref{eq:rr1}) into
\begin{eqnarray}
  &  & \delta \left( \frac{\tmmathbf{p}_f^2 -\tmmathbf{p}_i^2}{2} \right) 
  \frac{f \left( \tmmathbf{p}_f +\tmmathbf{q}_{\bot}, \tmmathbf{p_i}
  +\tmmathbf{q}_{\bot} \right)}{\sqrt{\sigma \left( \tmmathbf{p_i}
  +\tmmathbf{q}_{\bot} \right)  \left| \tmmathbf{p_i} +\tmmathbf{q}_{\bot}
  \right|}} \frac{f^{\ast} \left( \tmmathbf{p}_f -\tmmathbf{q}_{\bot},
  \tmmathbf{p_i} -\tmmathbf{q}_{\bot} \right)}{\sqrt{\sigma \left(
  \tmmathbf{p_i} -\tmmathbf{q}_{\bot} \right)  \left| \tmmathbf{p_i}
  -\tmmathbf{q}_{\bot} \right|}} g \left( \tmmathbf{q}_{\bot} \right) . 
  \label{eq:rr2}
\end{eqnarray}
Here, we denote, for given $\tmmathbf{q} \neq 0$ the parallel contribution of
a vector $\tmmathbf{p}$ by $\tmmathbf{p}_{\|\tmmathbf{q}} = \left(
\tmmathbf{p} \cdot \tmmathbf{q} \right) \tmmathbf{q}/ q^2$ and the
perpendicular one by $\tmmathbf{p}_{\bot \tmmathbf{q}}
=\tmmathbf{p}-\tmmathbf{p}_{\|\tmmathbf{q}}$. These projections manifestly
guarantee the conservation of energy in the two scattering amplitudes
individually. As discussed in {\cite{Hornberger2006b}}, this treatment leads
directly to
\begin{eqnarray}
  M_{\tmop{in}} \left( \tmmathbf{P}, \tmmathbf{P}' ; \tmmathbf{Q} \right) & =
  & \int_{\tmmathbf{Q}^{\bot}} \mathd \tmmathbf{K}L \left( \tmmathbf{K},
  \tmmathbf{P}-\tmmathbf{Q}; \tmmathbf{Q} \right) L^{\ast} \left(
  \tmmathbf{K}, \tmmathbf{P}' -\tmmathbf{Q}; \tmmathbf{Q} \right) 
  \label{eq:Min3}
\end{eqnarray}
with
\begin{eqnarray}
  L \left( \tmmathbf{K}, \tmmathbf{P}; \tmmathbf{Q} \right) & = &
  \sqrt{\frac{n_{\tmop{gas}} m}{Qm_{\ast}^2 }} \mu \left( \tmmathbf{K}_{\bot
  \tmmathbf{Q}} + \left( 1 + \frac{m}{M} \right) \frac{\tmmathbf{Q}}{2} +
  \frac{m}{M} \tmmathbf{P}_{\|\tmmathbf{Q}} \right)^{1 / 2} \nonumber\\
  &  & \times f \left( \tmop{rel} \left( \tmmathbf{K}_{\bot \tmmathbf{Q}},
  \tmmathbf{P}_{\bot \tmmathbf{Q}} \right) - \frac{\tmmathbf{Q}}{2},
  \tmop{rel} \left( \tmmathbf{K}_{\bot \tmmathbf{Q}}, \tmmathbf{P}_{\bot
  \tmmathbf{Q}} \right) + \frac{\tmmathbf{Q}}{2} \right) .  \label{eq:Fdef}
\end{eqnarray}
The integration in (\ref{eq:Min3}) is over the plane $\tmmathbf{Q}^{\bot} =
\left\{ \tmmathbf{K} \in \mathbbm{R}^3 : \tmmathbf{K} \cdot \tmmathbf{Q}= 0
\right\}$ perpendicular to the momentum transfer $\tmmathbf{Q}$. Clearly, the
function $L$ contains all the details of the collisional interaction with the
gas. It involves the elastic scattering amplitude $f \left(
\tmmathbf{p}_{\tmop{out}}, \tmmathbf{p}_{\tmop{in}} \right)$, the momentum
distribution function $\mu \left( \tmmathbf{p} \right)$ of the gas, and its
number density $n_{\tmop{gas}}$.

\subsection{Operator form of the quantum linear Boltzmann equation}

The particular form (\ref{eq:Dtrho2}) with (\ref{eq:Min3}) and (\ref{eq:Fdef})
admits the master equation to be written in a representation-independent
fashion. One finds the form {\cite{Hornberger2006b}}
\begin{eqnarray}
  \partial_t \rho & = & \frac{1}{\mathi \hbar} \left[ \mathsf{H}, \rho \right]
  + \bigintlim \mathd \tmmathbf{Q} \int_{\tmmathbf{Q}^{\bot}} \mathd
  \tmmathbf{K} \left\{ \mathsf{L} _{\tmmathbf{Q}, \tmmathbf{K}} \rho
  \mathsf{L} _{\tmmathbf{Q}, \tmmathbf{K}}^{\dag} - \frac{1}{2} \rho
  \mathsf{L} _{\tmmathbf{Q}, \tmmathbf{K}}^{\dag} \mathsf{L} _{\tmmathbf{Q},
  \tmmathbf{K}} - \frac{1}{2}  \mathsf{L} _{\tmmathbf{Q}, \tmmathbf{K}}^{\dag}
  \mathsf{L} _{\tmmathbf{Q}, \tmmathbf{K}} \rho \right\} \bignone 
  \label{eq:qlbe}
\end{eqnarray}
with the Lindblad operators $\mathsf{L} _{\tmmathbf{Q}, \tmmathbf{K}}$ defined
by
\begin{eqnarray}
  \mathsf{L} _{\tmmathbf{Q}, \tmmathbf{K}} & = & \mathe^{i \mathsf{X \cdot
  \tmmathbf{Q}/ \hbar}} L \left( \tmmathbf{K}, \mathsf{P} ; \tmmathbf{Q}
  \right) .  \label{eq:Ldef}
\end{eqnarray}
Here, the position and the momentum operator of the Brownian particle are
denoted by $\mathsf{X}$ and $\mathsf{P}$, respectively. Reassuringly, this
form of the master equation is in agreement with the general structure of a
translation-invariant and completely positive master equation, as
characterized by Holevo {\cite{Holevo1996a}}, see the discussion in
{\cite{Petruccione2005a,Vacchini2005a}}. Moreover, on the momentum diagonal it
reduces to the classical linear Boltzmann equation for the momentum
distribution function.

\begin{figure}[tb]
  \resizebox{2.5in}{!}{\epsfig{file=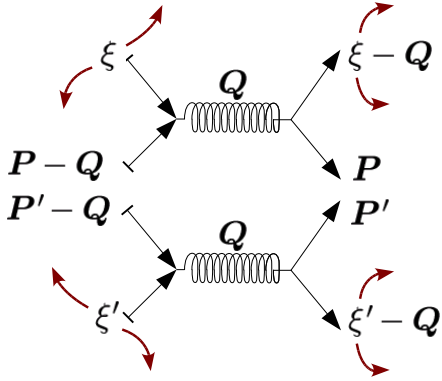}}
  \hspace{.2in}
  \begin{minipage}[b]{3.5in}
  \caption{A pictorial view of the pairs of ``two-particle trajectories''
  entering the in-rate (\ref{eq:Min3}) with (\ref{eq:Fdef}). For a given pair
  of final Brownian particle momenta, $\tmmathbf{P}$ and $\tmmathbf{P}'$, and
  for given common momentum exchange $\tmmathbf{Q}$, there are many different
  pairs of $\left( \xi, \xi' \right)$ of initial gas momenta which are
  consistent with the conservation of energy and total momentum. The integral
  in (\ref{eq:Min3}) covers them all and weights them according to their
  distribution in the gas.\label{fig3}}
  \end{minipage}
\end{figure}

Equation (\ref{eq:qlbe}) should be viewed as a full quantum version of the
linear Boltzmann equation. Although its representation independent operator
form (\ref{eq:Fdef})-(\ref{eq:Ldef}) might appear complicated at first sight,
it has in fact a rather suggestive interpretation when viewed in the momentum
representation (\ref{eq:Dtrho2}). To see this let us define as a possible
two-particle momentum trajectory $\left( \tmmathbf{P}_{\tmop{in}},
\tmmathbf{p}_{\tmop{in}} \right) \rightarrow \left( \tmmathbf{P}_{\tmop{out}},
\tmmathbf{p}_{\tmop{out}} \right)$ any combination of initial and final
momenta of the Brownian and the gas particle, respectively. The
{\emph{allowed}} two-particle trajectories are those which conserve the total
momentum end energy. The in-rate (\ref{eq:Min3}) in the momentum
representation (\ref{eq:Dtrho2}) can now be understood as the integral over
the scattering amplitudes of all allowed {\emph{pairs}} of two-particle
trajectories, which end at $\tmmathbf{P}$ and $\tmmathbf{P}'$, respectively,
and correspond to the common momentum exchange $\tmmathbf{Q}$, cf.~Fig
\ref{fig3}. The integration measure, i.e., the weight of the pairs of
trajectories, is simply given by the distribution in the gas, as restricted by
the choice of $\tmmathbf{P}$, $\tmmathbf{P}'$, and $\tmmathbf{Q}$, and by the
requirement of energy and momentum conservation.

\subsection{Relation to previous results}

A master equation very similar to the monitoring result
(\ref{eq:Fdef})-(\ref{eq:Ldef}) was proposed, on a slightly more heuristic
basis, in 1995 by Di\'osi {\cite{Diosi1995a}}. In the present notation his
equation is given by (\ref{eq:qlbe}) and (\ref{eq:Ldef}) provided the
functions $L \left( \tmmathbf{K}, \tmmathbf{P}; \tmmathbf{Q} \right)$ are
replaced by
\begin{eqnarray}
  L_{\text{Di\'osi}} \left( \tmmathbf{K}, \tmmathbf{P}; \tmmathbf{Q} \right) &
  = & \sqrt{\frac{n_{\tmop{gas}} m}{Qm_{\ast}^2}} \mu \left( \tmmathbf{K}+
  \left( 1 + \frac{m}{M} \right) \frac{\tmmathbf{Q}}{2} + \frac{m}{M}
  \tmmathbf{P} \right)^{1 / 2} f \left( \frac{m_{\ast}}{m} \tmmathbf{K}-
  \frac{\tmmathbf{Q}}{2}, \frac{m_{\ast}}{m} \tmmathbf{K}+
  \frac{\tmmathbf{Q}}{2} \right) . \nonumber\\
  &  &  \label{eq:FDiosi}
\end{eqnarray}
The main difference with respect to the present result (\ref{eq:Ldef}) is that
the arguments of the scattering amplitude do not depend on $\tmmathbf{P}$.
This implies that the complex rate function $M_{\tmop{in}}$ is determined by
the differential cross section $\mathd \sigma / \mathd \Omega = \left| f
\right|^2$, rather the individual complex scattering amplitudes of the
collision trajectories. This indicates that it does not incorporate the
non-parallel pairs of trajectories, $\xi \neq \xi'$, though, a priori, one
would expect them to contribute to the rate as well. On the diagonal, i.e.,
for $\tmmathbf{P}=\tmmathbf{P}'$, Di\'osi's equation coincides with the
present result (\ref{eq:Fdef})-(\ref{eq:Ldef}); however they {\emph{differ}}
in the ``Brownian limit'' of small momentum transfers, thus predicting
different coefficients for the corresponding Fokker-Planck equation.

Related and more recent works can be found in
{\cite{Altenmuller1997a,Dodd2003a}}. The most relevant development for our
case is the theory by Vacchini {\cite{Vacchini2000a,Vacchini2001a}}, which
uses the van Hove expression for macroscopic scattering to relate the
two-point correlation function of the gas with the collision kernel, a
treatment which is valid in the weak-coupling limit. It turns out that his
equation coincides with the limiting form of the present result if one
replaces the scattering amplitude in (\ref{eq:Fdef}) by its Born approximation
$f_{\text{B}}$. The reason is that $f_{\text{B}}$ depends only on the
difference of the momenta so that the $\tmmathbf{P}$- and
$\tmmathbf{K}$-dependence of the scattering amplitude vanishes in this limit.
Incidentally, this shows that the non-perturbative form of the master equation
cannot be specified by a weak coupling calculation alone. In particular, it is
not obtained by simply replacing the Born approximation of the scattering
amplitude with the exact one.

Another limiting form of the master equation is the case of an infinitely
massive Brownian particle, $m / M \rightarrow 0$, where it describes no
dissipation, but pure spatial decoherence. As one expects, the present result
(\ref{eq:Fdef})-(\ref{eq:Ldef}) reduces in this limit to the proper master
equation for collisional decoherence {\cite{Hornberger2003b}}, which was
recently tested experimentally {\cite{Hornberger2003a,Hornberger2004a}}.

\section{Conclusions}

In summary, I presented a general method of incorporating proper
time-dependent scattering theory into the dynamic description of open quantum
systems. Its derivation is based on the theory of generalized and continuous
measurements and it yields completely positive master equations, which account
for the environmental interaction in a non-perturbative fashion. When applied
to either the case of an immobile, complex object or the case of a Brownian
point particle, it provides a detailed and realistic account of the
dissipation and decoherence effects induced by the presence of a gaseous
environment.

{\subsection*{Acknowledgments}}

I would like to thank the organizers of DICE2006, and in particular
Hans-Thomas Elze, for bringing together researchers from rather different
fields of physics, and thus creating a very stimulating atmosphere. I also
thank Bassano Vacchini for helpful discussions. This work was supported by the
DFG Emmy Noether program.

\vspace*{\baselineskip}

\end{document}